\documentclass[aps,prl,twocolumn,superscriptaddress,floatfix]{revtex4}

\pdfoutput=1
\usepackage{soul,xcolor}
\usepackage{bm}
\usepackage{amsmath}
\usepackage{amssymb}
\usepackage{graphicx}
\usepackage{color}
\usepackage{xcolor}\definecolor{ao}{rgb}{0.0,0.0,1.0}

\definecolor{br}{rgb}{1.0, 0.22, 0.0}

\def\sig{{\mbox{\boldmath{$\sigma$}}}}
\input{epsf}
\begin{document}

\setstcolor{red}

\title{Photo-spintronics of spin-orbit active electric weak links}

\author{R. I. Shekhter}
\affiliation{Department of Physics, University of Gothenburg, SE-412
96 G{\" o}teborg, Sweden}

\author{O. Entin-Wohlman}
\affiliation{Raymond and Beverly Sackler School of Physics and Astronomy, Tel Aviv University, Tel Aviv 69978, Israel}
\affiliation{Physics Department, Ben Gurion University, Beer Sheva 84105, Israel}


\author{M. Jonson}
\email{mats.jonson@physics.gu.se}
\affiliation{Department of Physics, University of Gothenburg, SE-412
96 G{\" o}teborg, Sweden} 
\affiliation{SUPA, Institute of Photonics
and Quantum Sciences, Heriot-Watt University, Edinburgh, EH14 4AS,
Scotland, UK}

\author{A. Aharony}
\affiliation{Raymond and Beverly Sackler School of Physics and Astronomy, Tel Aviv University, Tel Aviv 69978, Israel}
\affiliation{Physics Department, Ben Gurion University, Beer Sheva 84105, Israel}

\date{\today}

\begin{abstract}
We show that a carbon nanotube can serve as a functional electric weak link performing 
photo-spintronic transduction. A spin current, facilitated by strong spin-orbit interactions
in the nanotube and not accompanied by a charge current, is induced in a 
device containing the nanotube weak link by 
circularly polarized microwaves. Nanomechanical tuning of the photo-spintronic transduction can 
be achieved due to the sensitivity of the spin-orbit interaction to geometrical deformations of the 
weak link.
\end{abstract}


\maketitle

\section{Introduction}

Spintronics is a rapidly developing research area of modern solid state physics. In contrast to 
traditional electronics, where the electric charge of electrons is in focus, the field of spintronics relies on 
another fundamental property of electrons, {\em viz.} their magnetic moment, which is associated 
with their spin degree of freedom. 

Issues related to the electrical control of spin currents as well as to spin control of charge currents 
are at the heart of spintronics research today, both from a fundamental and an applied perspective 
\cite{review}. Recently it has been suggested that such controls may be effectively implemented in 
nanodevices containing an electric weak link with strong spin-orbit interactions (SOI) that bridges 
bulk electrodes \cite{Shekhter.2013.2014x, Shekhter.2016x, Shekhter.2017}. 
In \cite{Shekhter.2013.2014x} it was demonstrated that a spin-orbit coupling results in a ``splitting" 
of the spin of electrons passing through such a weak link (Rashba spin splitting), which under certain 
conditions may generate a spin current. This was shown to occur if an imbalance of the population of 
spin states in the electrodes is established by spin-flip assisted electronic transitions due to the 
absorption (or emission) of circularly polarized photons created by microwave pumping
\cite{microwaves}. 
The SOI-induced spin generation inside the weak link makes it a point-like source of a spin current
due to a photo-spintronic effect on the nanometer length scale.
Estimations show, however, that if the SOI is caused by an external electric field, as implicitly assumed
in \cite{Shekhter.2013.2014x}, that field has to be quite strong for the induced
Rashba spin splitting to be significant. The aim of the present work is to demonstrate that a much 
stronger photo-spintronic transduction effect can be achieved if a material with an intrinsic SOI, here
assumed to be induced by stresses, is used for the weak link.

The precise form of the stress-induced SOI depends on the material used for the electric weak link and 
the type of strains involved. In a single-wall carbon nanotube, which will be considered here, the strain 
can be thought of as occurring when a graphene ribbon is rolled up to form a tube. The strain-induced 
SOI in a simple one-dimensional model of such a nanotube is described by the Hamiltonian 
\begin{align}
\hat {\cal H}^{\rm strain}_{\rm so} = \hbar v_{\rm F }^{}k^{\rm strain}_{\rm so} \sig \cdot \hat{\bf n}\,,
\label{HAMstrain}
\end{align} 
where $v_{\rm F}$ is the Fermi velocity, $k^{\rm strain}_{\rm so}$ is a phenomenological parameter that 
gives the strength of the SOI in units of inverse length, $\sig$ is a vector whose components are the Pauli 
matrices $\sigma_{x,y,z}$, and $\hat{\bf n}$ is a unit vector pointing along the longitudinal axis of the 
nanotube in the direction of electron propagation ($\hat{\bf n}=\hat{\bf k}$). 
Equation (\ref{HAMstrain}), which was used in Ref. \onlinecite{Shekhter.2016x}, is a simplified 
form of the SOI Hamiltonian previously derived for a realistic model of such a nanotube \cite{Rudner.Rashba.2010}. 

The SOI active weak-link device shown in Fig.~1 comprises a nanowire that 
bridges two bulk electronic reservoirs. The spin-orbit interaction described by Eq.~(\ref{HAMstrain}) is restricted 
to the nanowire and has the effect of scattering the spins of electrons that pass through the wire. 
Following the approach developed in Refs.~\onlinecite{Shekhter.2013.2014x, Shekhter.2016x, Shekhter.2017} we 
will describe the transfer of electrons through the nanowire-based weak link with the help of a spin-dependent tunnel 
Hamiltonian. Hence, the total Hamiltonian of the system can be written as a sum of three parts, 
\begin{align}
\hat {\cal H} = \hat {\cal H}_L + \hat {\cal H}_R  + \hat {\cal H}_T  \,,
\label{HAM}
\end{align}
where 
\begin{align}
\hat {\cal H}_{L(R)} = \sum_{{\bf k} ({\bf p})\sigma} \varepsilon^{}_{k(p)}c^\dag_{{\bf k}({\bf p})\sigma}c^{}_{{\bf k}({\bf p})\sigma}
\label{HAM_LR}
\end{align}
are Hamiltonians that describe the electrons in the left and right leads. These electrons are characterized by 
the momentum quantum numbers ${\bf k}$ and ${\bf p}$, respectively, and by the spin projection on the $\hat{\bf z}-$axis. We label 
the latter by $\sigma=\pm 1$, so that the spin projections are $s=\hbar\sigma/2$. The tunnel Hamiltonian is 
expressed in terms of the probability amplitudes $[W^{}_{{\bf p},{\bf k}}]^{}_{\sigma,\sigma'}$ for electron 
transmission through the wire,
\begin{align}
\hat {\cal H}^{}_{T}=\sum_{{\bf k},{\bf p}}\sum_{\sigma,\sigma '}(c^{\dagger}_{{\bf p}\sigma '}[W^{}_{{\bf p},{\bf k}}]^{}_{\sigma',\sigma}c^{}_{{\bf k}\sigma}+{\rm H.c.})\ .
\label{HT}
\end{align}
These amplitudes have to be calculated taking the spin dynamics given by 
Hamiltonian (\ref{HAMstrain}) into account.

\begin{figure}[htp]
\vspace{-1cm}
\includegraphics[width=9cm]{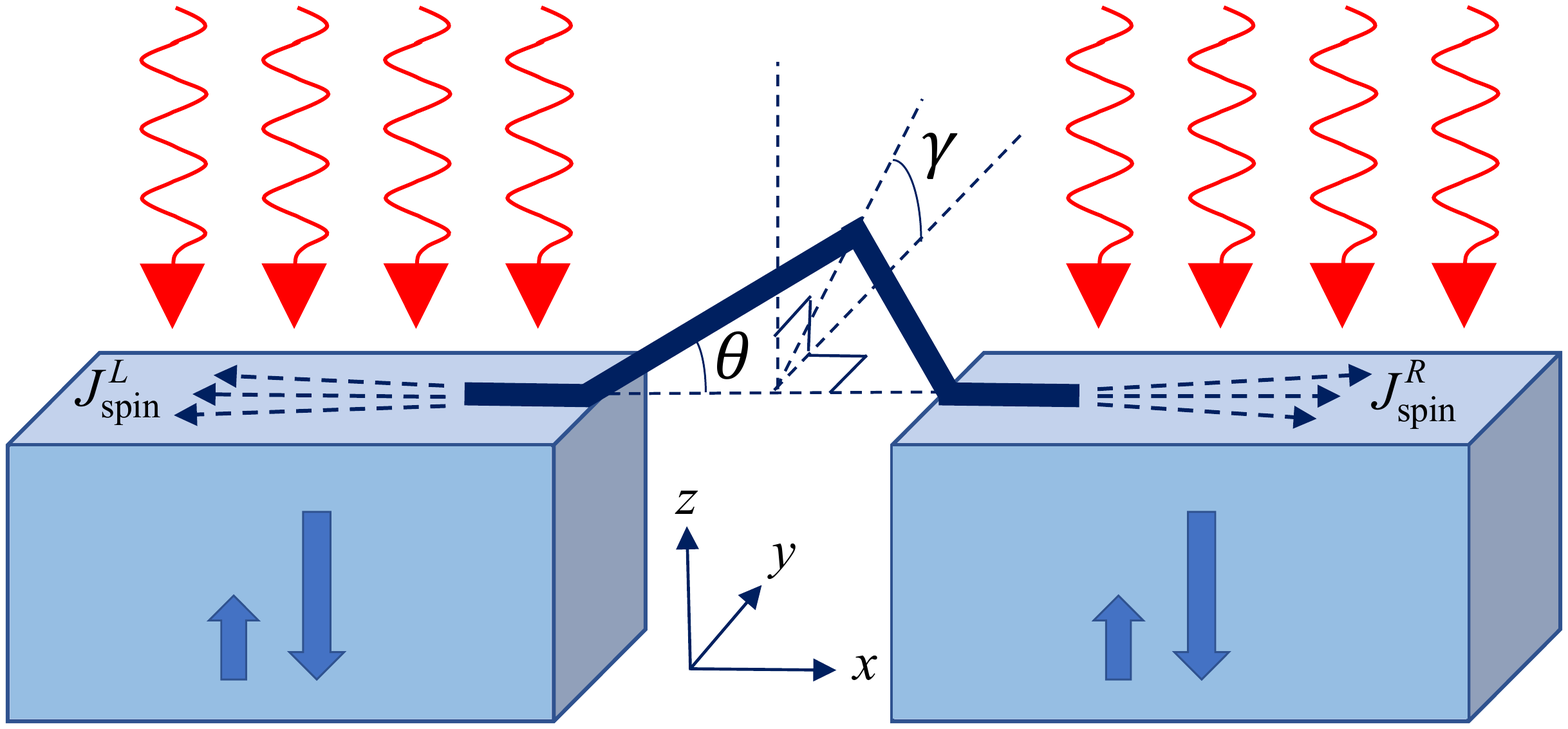}
\vspace{-1.5cm}
\caption{Schematic picture of the device considered. A bent nanowire with a strong spin-orbit interaction bridges 
two bulk reservoirs. The bent nanowire is modeled by two equal-length straight wire segments which form angles 
$\theta$ and $-\theta$ with the $\hat{\bf x}-$axis, respectively, while the plane that contains the nanowire is tilted by an angle
$\gamma$ away from the $\hat{\bf y}-$axis towards the $\hat{\bf z}-$axis.  The device is irradiated by circularly-polarized microwave
radiation (wavy arrows) that creates  a difference in the population of spin-up and spin-down
electrons (thick vertical arrows) corresponding to a ``spin biasing" of the device. Spin-flip transitions induced by the 
spin-orbit interaction in the nanowire create a spin current $J_{\rm spin} = J^L_{\rm spin} + J^R_{\rm spin}$
traveling out from the wire with a magnitude that depends 
on the angles $\theta$ and $\gamma$. 
}
\label{Fig1}
\end{figure}

\section{Spin-biased electric weak link}

Electronic transport through the weak link shown in Fig.~\ref{Fig1} can be induced in a number of different ways. 
The standard method would be to apply a voltage bias $V$ across the link, so that the chemical potentials for 
electrons in the left and right reservoirs are shifted by a small amount $eV$ with respect to each other.
The result of such {\em charge biasing} is that excess charge of opposite polarity is accumulated on either side 
of the link, which leads to an electrical current through the link in a direction that counters the
bias-induced charge imbalance. 

Another method, which will be in focus here and which is illustrated in Fig.~1, is to arrange for the 
chemical potential to be different for the two possible projections of the electron spin along a certain axis. Such 
{\em spin biasing} can be achieved by illuminating the entire device with circularly-polarized microwave radiation 
of a frequency that enables electron spin-flip assisted photon absorption. This photonic pumping of the 
electronic spin creates an imbalance between the number of electrons with opposite spin projections on an axis
defined by the direction of the radiation and leads to a spin-dependent shift in the chemical potentials for 
electrons with opposite spin projections. The magnitude of the shift, which extends throughout the device, 
depends on the intensity of the radiation and on the spin relaxation rate. 
If we assume that the SOI, which is restricted to the weak link, is the dominant spin relaxation mechanism 
the SOI becomes a source of a spin current, which flows out of the weak link into the two reservoirs and 
which counteracts the spin pumping.
Referring to Fig.~\ref{Fig1}, we note that both the orientation of the plane that contains the weak link with respect to the 
pumped spin orientation and the bending angle of the link can be used to 
tune the generated spin current.

The spin bias $U$ can be defined by noting that for $\hat {\cal H}_T = 0$ the electron spin reservoirs are described 
by Fermi-Dirac distributions with different  chemical potentials, 
\begin{align}
f^{\sigma}_{{\bf k}({\bf p})}=\langle  c^{\dagger}_{{\bf k}({\bf p})\sigma}c^{}_{{\bf k}({\bf p})\sigma}\rangle ^{}_{\hat{\cal H}^{}_{T}=0}=n^{}_{\rm F}(\varepsilon^{}_{k(p)}-\mu^{}_{\sigma})\ ,
\end{align}
where
\begin{align}
\mu^{}_{\sigma}=\mu-\sigma U/2\ ,\ \ \sigma=\pm 1\ ,
\end{align}
and $\mu$ is the  chemical potential of both leads at equilibrium.
The spin current generated by electrons tunneling out of the SOI-active weak link can be obtained as
the time derivative of the total spin, $\langle \hat{S}\rangle $, of the electrons, 
\begin{align}
J^{}_{\rm spin} = \frac{d \langle\hat{S}\rangle }{d t} = \sum_{\sigma = \pm 1}
\frac{\hbar\sigma}{2}
\langle\frac{ d \hat {\cal N}^{}_\sigma}{d t}\rangle\ ,
\label{spincurrent}
\end{align}
where 
\begin{align}
\hat {\cal N}^{}_\sigma = \hat {\cal N}^{}_{L\sigma} + \hat {\cal N}^{}_{R\sigma}\ ,\ \ 
\hat {\cal N}^{}_{L(R)\sigma} = \sum_{{\bf k}({\bf p})} c^{\dagger}_{{\bf k}({\bf p})\sigma}c^{}_{{\bf k}({\bf p})\sigma}\ .
\end{align}
A straightforward calculation of the spin current (\ref{spincurrent}) can be done using the tunnel Hamiltonian (\ref{HAMstrain}) 
to lowest order in perturbation theory \cite{Shekhter.2013.2014x}. Then, one finds for the spin conductance $G_{\rm spin}$  (defined in analogy
with the electrical conductance $G$) the relation
\begin{align}
G^{}_{\rm spin} = J^{}_{\rm spin}/U \quad {\rm for} \quad U \to 0 \ .
\label{gspin}
\end{align}

\section{Rashba spin splitting as a source of spin generation in SOI active weak link}

In order to calculate the spin current given by Eq.~(\ref{spincurrent}) one needs to evaluate the electronic transmission 
amplitudes $[W^{}_{{\bf p},{\bf k}}]^{}_{\sigma,\sigma'}$, which appear in the tunnel Hamiltonian (\ref{HT}), 
and specifically their dependence on the spin-orbit interaction 
as given by Eq.~(\ref{HAMstrain}).
In our simple model the weak link consists of two straight parts of equal length $\vert {\bf R}_L\vert = \vert {\bf R}_R\vert =d/2$
joined by a bend. 
Neglecting the momentum (but not the spin) dependence, the probability amplitude for an electron 
of energy $E$ to pass from, say, the left to the right lead can be written as a product of five factors,
\begin{align}
W = T^{}_R \,G({\bf R}^{}_R; E)\, {\cal T}\, G({\bf R}^{}_L; E)\,T^{}_L \ .
\label{5factors}
\end{align}
Here $W$ is a 2$\times$2 matrix in spin space, $T_{L(R)}$ is the probability amplitude to tunnel from the wire 
to the left (right) lead and ${\cal T}$ is the transfer matrix through the bend 
in the wire. 
In Ref. \onlinecite{Shekhter.2017} the Green's function $G({\bf R}_{L(R)}; E)$ for the straight segments of the wire, 
in which the SOI interaction takes place, was evaluated  for a Hamiltonian of the form
\begin{align}
\hat{\cal H} = \frac{\hbar^2 k^2}{2m^*} + {\bf Q}({\bf k}) \cdot \sig\, .  
\end{align}
A comparison with Eq.~(\ref{HAMstrain}) shows that in the present case 
$ {\bf Q}({\bf k}) = \hbar v_{\rm F }^{}k^{\rm strain}_{\rm so}\hat{\bf n}$. Hence, from Eqs. (A12) and (A13) of
Ref. \onlinecite{Shekhter.2017} we conclude that
\begin{align}
G({\bf R}^{}_{L(R)}; E) &= G_0({\bf R}^{}_{L(R)}; E)\nonumber\\
&\times [ \cos(\alpha) - i \sin(\alpha) \hat {\bf n}^{}_{L(R)} \cdot \sig]
\ ,
\end{align}
where we have used the short-hand notation $k_{\rm so}^{\rm strain}d/2 \equiv \alpha$, which
is a measure of the strength of the SOI, and where
\begin{align}
G_0({\bf R}^{}_{L(R)}; E) = i\pi (m^*/\hbar^2 k^{}_0)\exp[i k^{}_0 | {\bf R}_{L(R)} |]\ , 
\end{align}
with $k_0=(2m^*E/\hbar^2)^{1/2}$, is the propagator on the left (right) segment in the absence of SOI.
It follows that we can factor out the dependence on the SOI and write the amplitude given by Eq.~(\ref{5factors}) as
\begin{align}
W = W^{}_0 {\cal W} \, ,
\end{align}
where
\begin{align}
W^{}_0 = T^{}_R \,G^{}_0(|{\bf R}^{}_R|; E)\, {\cal T} \,G^{}_0(|{\bf R}^{}_L|; E)\,T^{}_L
\end{align}
is the SOI-independent part and 
\begin{align}
{\cal W} = [ \cos(\alpha) -i \sin(\alpha) \hat {\bf n}^{}_R \cdot \sig ] [ \cos(\alpha) -i \sin(\alpha) \hat{\bf  n}^{}_L \cdot \sig ]
\label{calW}
\end{align}
contains the effect of the SOI.

\begin{figure}[htp]
\vspace{-1cm}
\includegraphics[width=10cm]{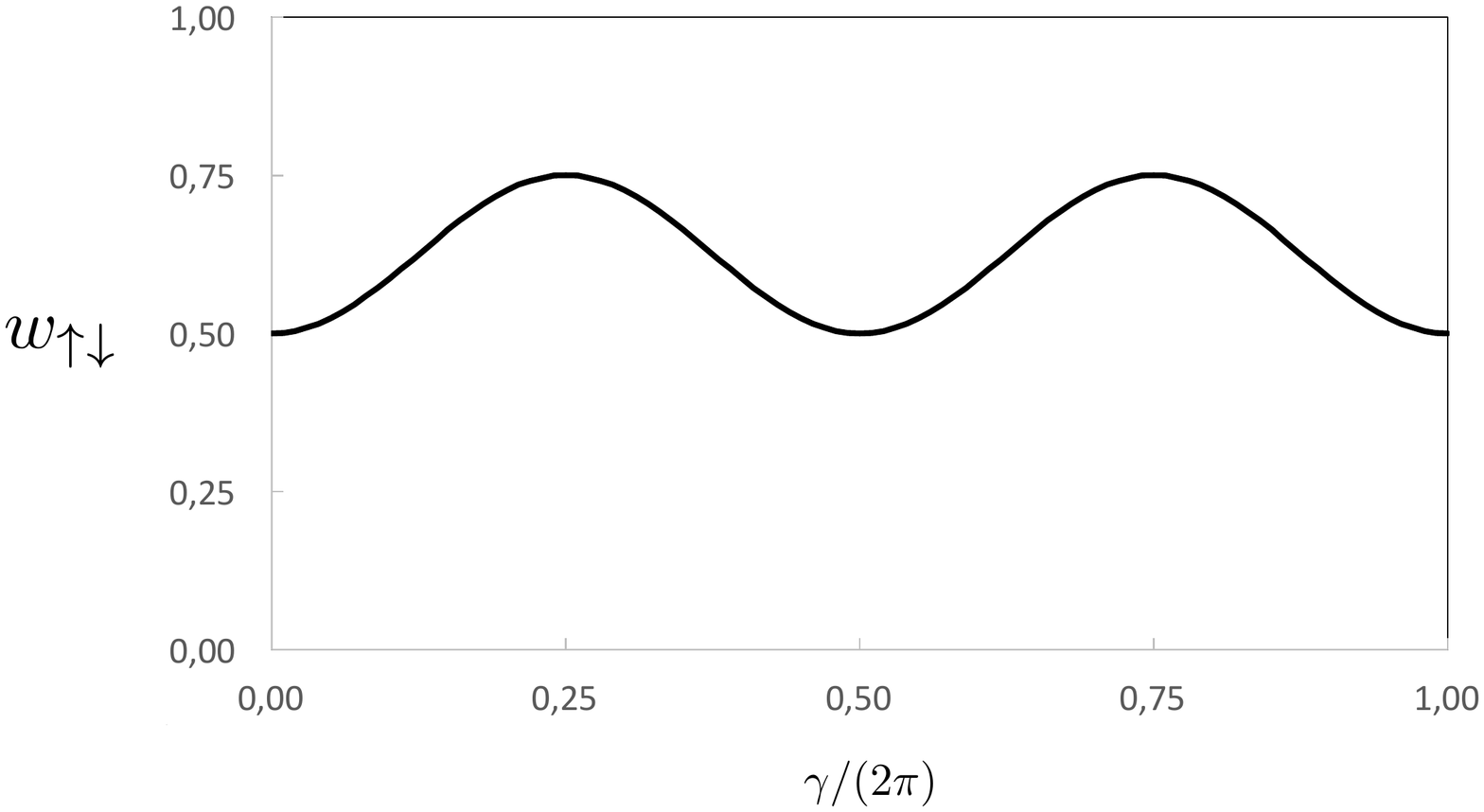}
\vspace{-1.5cm}
\caption{The dimensionless spin conductance $w_{\uparrow\downarrow}(\theta,\gamma)$ defined in Eqs.  (\ref{wupdown}) and (\ref{gspin}), 
plotted as a function of the tilt angle $\gamma$ of the plane that contains the nanowire weak
link (see Fig.~\ref{Fig1}). The spin conductance oscillates between a maximal value $w_{\rm max}^{\alpha}(\theta)$
and a minimal value $w_{\rm min}^{\alpha}(\theta)$, which both depend on the bend 
angle $\theta$
and the strength $\alpha$ of the spin-orbit interaction in the wire. Here $\alpha=\theta=\pi/4$ for which
$w_{\rm max}^{\alpha}(\theta)=0.75$ and $w_{\rm min}^{\alpha}(\theta)=0.50\,$.
}
\label{Fig2}
\end{figure}

\begin{figure}[htp]
\vspace{-1cm}
\includegraphics[width=10cm]{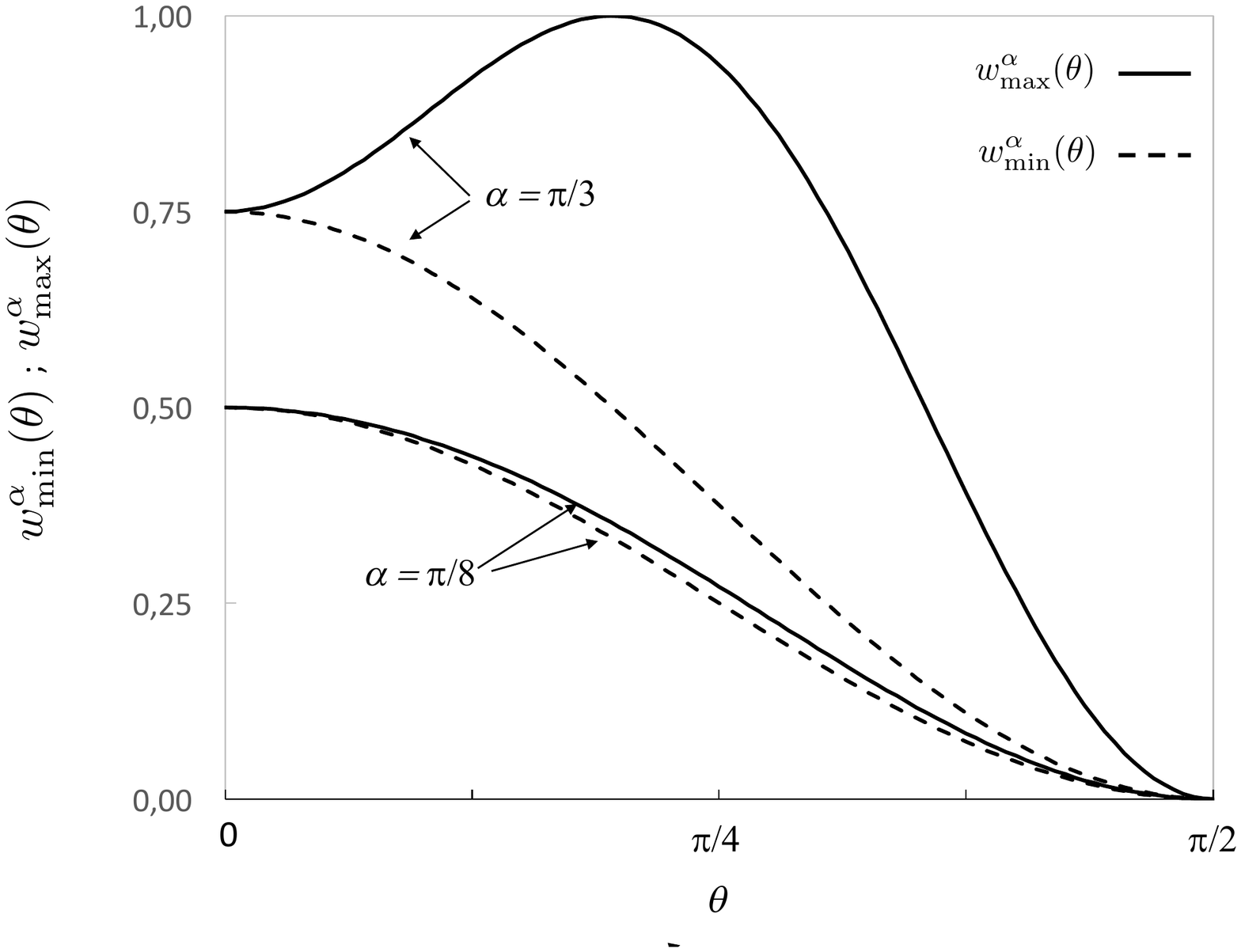}
\vspace{-1.5cm}
\caption{The maximal value $w_{\rm max}^{\alpha}(\theta)$ of the dimensionless spin conductance 
$w_{\uparrow\downarrow}(\theta,\gamma)$ [obtained for $\sin^2(\gamma)=1$] and the minimal value  
$w_{\rm min}^{\alpha}(\theta)$ of $w_{\uparrow\downarrow}(\theta,\gamma)$ [obtained for $\sin^2(\gamma)=0$]
plotted as functions of the nanowire bend angle $\theta$ defined in Fig.~\ref{Fig1}.
The two pairs of curves are for two different values of the spin-orbit interaction strength $\alpha$ in the 
nanowire: the upper  pair pertains  for $\alpha=\pi/3$, for which $\sin^{2}\alpha >1/2$, and the lower one 
pertains for $\alpha=\pi/8$, for which $\sin^2(\alpha) < 1/2$.
Note that when $\sin^2(\alpha) \geq 1/2$ the maximal 
spin conductance  $w_{\rm max}^{\alpha}(\theta)$ reaches the value 1 
for $\cos^2(\theta) = 1/[2\sin^2(\alpha)]$.
When $\sin^2(\alpha) < 1/2$ both $w_{\rm max}^{\alpha}(\theta)$ and $w_{\rm min}^{\alpha}(\theta)$ decrease monotonically 
from $\sin^2(2\alpha)$ to 0.
}
\label{Fig3}
\end{figure}

In our geometry the $\hat{\bf z}-$axis  is the spin quantization axis, and the bent  
wire lies in a plane that (i) contains the $\hat{\bf x}-$axis
and (ii) is rotated by an angle $\gamma$ away from the $\hat{\bf y}-$axis towards the $\hat{\bf z}-$axis. In the plane of the wire, the left (right) straight 
leg of the wire forms an angle $\theta$ (-$\theta$) with the $\hat{\bf x}-$axis. In other words,
\begin{align}
\hat{\bf n}^{}_L &= \cos(\theta) \hat{\bf  x} + \sin(\theta)[ \cos(\gamma) \hat {\bf y} + \sin(\gamma) \hat {\bf z}]\ ,\nonumber \\
\hat {\bf n}^{}_R &= \cos(\theta) \hat {\bf x} - \sin(\theta)[ \cos(\gamma) \hat{\bf  y} + \sin(\gamma) \hat {\bf z} ] \ ,
 \end{align}
which means that we can write Eq.~(\ref{calW}) in a matrix form,
\begin{align}
{\cal W} = A -i{\bf B} \cdot \sig\ ,
\end{align} 
where
\begin{align}
A&=\cos^{2}_{}(\alpha)-\sin^{2}_{}(\alpha)\hat{\bf n}^{}_{R}\cdot\hat{\bf n}^{}_{L}\nonumber\\
&= \cos^2(\alpha) - \sin^2(\alpha) \cos(2\theta)\ ,
\end{align}
and
\begin{align}
{\bf B}&=\sin(\alpha)\cos(\alpha)(\hat{\bf n}_{R}+\hat{\bf n}_{L}^{})+\sin^{2}(\alpha)\hat{\bf n}^{}_{R}\times\hat{\bf n}^{}_{L}\nonumber\\
&=\hat{\bf x}\sin(2\alpha)\cos(\theta)-\hat{\bf y}\sin^2(\alpha)\sin(2\theta)\sin(\gamma)\nonumber\\
&+\hat{\bf z} \sin^2(\alpha)\sin(2\theta)\cos(\gamma)\ .
\end{align}
Hence, the probability for a SOI-induced spin-flip transition is
\begin{align}
\label{wupdown}
w^{}_{\uparrow\downarrow} &\equiv | {\cal W}^{}_{\uparrow\downarrow} |^{2} = |B^{}_x |^{2}_{} + |B^{}_y |^{2}_{}\nonumber\\
&= \sin^2(2\alpha)\cos^2(\theta) + \sin^4(\alpha)\sin^2(2\theta)\sin^2(\gamma) \ .
\end{align}
The spin conductance can now be expressed in terms of the spin-flip probability $w_{\uparrow\downarrow}$
as 
\begin{align}
G_{\rm spin} = \frac{G}{e^2/\hbar} \,w_{\uparrow\downarrow}(\theta,\gamma)\, .
\label{gspin}
\end{align}
Equations (\ref{wupdown}) and (\ref{gspin}) represent the main results of the paper. The strength of the SOI, 
characterized by the dimensionless parameter $\alpha=k_{\rm so}^{\rm strain} d/2$, determines the 
amount of photo-spintronic transduction that can be achieved  by the studied Rashba spin-splitter device. 
The sensitivity of the effect to the geometry of the experimental set-up opens the possibility  for 
tuning the device nanomechanically, by varying the angles $\gamma$ and $\theta$. 
The dependence of the spin conductance on these experimentally accessible device
 parameters is illustrated 
in Figs.~\ref{Fig2} and \ref{Fig3}.

The photo-spintronic transduction occurs even for  a straight wire (i.e., when  $\theta=0$).
However, wire deformation provides a tool for a nanomechanical control of the generated spin current. Depending on 
the strength of SOI coupling $\alpha$, both a monotonic and a non-monotonic dependence of the dimensionless 
spin-conductance  $w_{\uparrow\downarrow}$ on mechanical deformations  can be achieved. 

Figure  \ref{Fig2}  illustrates that the dimensionless conductance, $w_{\uparrow\downarrow}$, is an oscillatory 
function of the angle $\gamma$. As is clear  from Eq.~(\ref{wupdown}),  the maximal [minimal] value,  
$w_{\rm max}^{\alpha}(\theta)$ [$w_{\rm min}^{\alpha}(\theta)$], of $w_{\uparrow\downarrow}=w_{\uparrow\downarrow}(\gamma)$ 
is achieved when the normal to the plane containing the nanowire weak link is perpendicular [parallel] to the spin quantization axis, i.e. when $\gamma = \pm \pi/2$ 
[$\gamma = 0$ or $\pi$] and therefore $\sin^2(\gamma)=1$ [$\sin^2(\gamma)=0$], independent of the values of $\alpha$ and $\theta$. 

For a strong enough SOI, i.e., when $\sin^2(\alpha) \ge 1/2$, the largest possible value,
$w_{\rm max}^{\alpha}(\theta)=1$,  is reached when the bending angle is $\theta=\theta_\alpha$, where 
$\cos^2(\theta_\alpha) = 1/[2\sin^2(\alpha)]$, as illustrated in Fig. \ref{Fig3}. On the other hand, for  $\sin^2(\alpha) < 1/2$ 
the effect is smaller since then $w_{\rm max}^{\alpha}(\theta) < \sin^2(2\alpha) < 1$.


\section{Conclusions}
In this paper we have shown that a nanowire, in which the electrons are subjected  to a spin-orbit interaction (SOI),
can be used as a functional electric weak link between SOI-inactive leads and serve as an essentially point-like 
source of a spin-current induced by circularly-polarized  microwave radiation. This spin current is not
accompanied by a charge current. The possibility to concentrate such ``pure"
spin-currents at the nanometer length scale suggests novel spintronic 
devices. 

Whether realistic applications are feasible crucially depends on how 
strong  a photo-spintronic effect can be realized in practice, which in turn depends on the strength of the SOI 
that can be achieved. Consider, for instance, a single-wall carbon nanotube.
Its  strain-induced SO energy gap $\Delta^{\rm strain}_{\rm so}$ has been measured to be around 0.4 meV  \cite{McEuen.2008}.
Since $\Delta^{\rm strain}_{\rm so}=2\hbar v_{\rm F}k^{\rm strain}_{\rm so}$, this value corresponds
to  $k^{\rm strain}_{\rm so}\approx 0.4\times 10^6$ m$^{-1}$ for $v_{\rm F}\approx 0.5\times 10^6$ m/s.
For nanowire lengths
$d$ of the order of a $\mu$m,
 $k^{\rm strain}_{\rm so}  d$ can therefore be of order $1$, which is large enough to allow 
the dimensionless spin conductance to be tuned near to its maximal value $w_{\uparrow\downarrow}=1$.

\begin{acknowledgments}
This work was partially supported by the Swedish Research
Council (VR), by the Israel Science Foundation
(ISF) and by the infrastructure program of Israel
Ministry of Science and Technology under contract
3-11173.
 \end{acknowledgments}

\end{document}